\vspace*{0.3truecm}
\documentclass[prb,aps,amsmath,amsfonts,amssymb,twocolumn, superscriptaddress,footinbib,showpacs, 10pt]{revtex4}
\usepackage{color}
\usepackage{amsmath}
\usepackage{graphicx}
\usepackage{subfigure}
\usepackage{epsfig}
\usepackage{soul}
\usepackage{ccaption}
\usepackage[justification=raggedright]{caption}

\newcommand{\beq}{\begin{equation}}
\newcommand{\beqn}{\begin{equation*}}
\newcommand{\eeqn}{\end{equation*}}
\newcommand{\eeq}{\end{equation}}
\newcommand{\barray}{\begin{eqnarray}}
\newcommand{\barrayn}{\begin{eqnarray*}}
\newcommand{\earray}{\end{eqnarray}}
\newcommand{\earrayn}{\end{eqnarray*}}
\newcommand{\balign}{\begin{align}}
\newcommand{\ealign}{\end{align}}
\newcommand{\ban}{\begin{align*}}
\newcommand{\ean}{\end{align*}}

\newcommand*{\Ham}{{\cal H}}
\newcommand*{\kkp}{kk^\prime}
\newcommand*{\be}{\begin{equation}}
\newcommand*{\ee}{\end{equation}}
\newcommand*{\bea}{\begin{eqnarray}}
\newcommand*{\eea}{\end{eqnarray}}

\newcommand{\cecoin}{CeCoIn$_5$}

\begin{document}

\title{Point-contact spectroscopy in heavy-fermion superconductors}
\author{Mikael Fogelstr\"om} 
\affiliation{Department of Microtechnology and Nanoscience, Chalmers, S-412 96 G\"oteborg, Sweden}
\author{W. K. Park}
\affiliation{Department of Physics and the Frederick Seitz Materials Research Laboratory, 
University of Illinois at Urbana-Champaign, Urbana, Illinois 61801-3080, USA}
\author{L. H. Greene}
\affiliation{Department of Physics and the Frederick Seitz Materials Research Laboratory, 
University of Illinois at Urbana-Champaign, Urbana, Illinois 61801-3080, USA}
\author{G. Goll}
\affiliation{DFG-Center for Functional Nanostructures, Karlsruhe Institute of Technology, D-76131 Karlsruhe, Germany}
\author{Matthias J. Graf}
\affiliation{Theoretical Division, Los Alamos National Laboratory, Los Alamos, New Mexico 87545, USA}

\date{\today}

\begin{abstract}
We develop a microscopic model to calculate point-contact spectra between a 
metallic tip and a superconducting heavy-fermion system.
We apply our tunneling model to the heavy fermion \cecoin, both in the normal and superconducting state. 
In point-contact and scanning tunneling spectroscopy many heavy-fermion materials, like \cecoin,
exhibit an asymmetric differential conductance,
$dI/dV$, combined with a strongly suppressed Andreev reflection signal in the superconducting state.
We argue that both features may be explained in terms of a multichannel tunneling model
in the presence of localized states near the interface. 
We find that it is not sufficient to tunnel into two itinerant bands of light and heavy electrons to explain the Fano line shape 
of the differential conductance. Localized states in the bulk or near the interface are an essential component 
for quantum interference to occur when an electron tunnels from the metallic tip of the point contact into the heavy-fermion system.
\end{abstract}

\pacs{74.55.+v, 74.70.Tx, 85.30.Hi}
%single particle tunneling (superconductivity), 74.55.+v
%heavy-fermion materials, 74.70.Tx
%point contact devices, 85.30.Hi

\maketitle
\section{Introduction}

Point-contact (PCS) and scanning tunneling spectroscopies (STS)
have been widely used to characterize the electronic behavior of 
heavy-electron materials, especially the transition into the superconducting state and the opening of an excitation gap.
Unlike point-contact junctions with conventional metals, most heavy-fermion materials show asymmetric conductances, which have
been difficult to explain.
In addition to asymmetric normal-state conductances, many heavy-fermion superconductors (HFS) are known for strongly suppressed
Andreev reflection signals. \cite{Nowack1987,DeWilde1994,Naidyuk1996,naidyuk1998}  
This is especially true for the heavy-fermion superconductor \cecoin. \cite{park2005,goll2005,goll2006,park2008a,park2008b,park09}
Hallmarks of superconductivity are phase coherence and Andreev reflection (AR), which require the existence 
of a condensate of Cooper pairs.
The AR signature occurs when a quasiparticle retro-reflects off a normal-superconducting (N/S) interface as a quasihole, while 
momentum and charge get carried across the interface by the Cooper pair.

The Blonder-Tinkham-Klapwijk (BTK) formulation describes the differential conductance in 
conventional N/S junctions remarkably well by
invoking a dimensionless barrier strength parameter $Z$, which depends on the barrier potential and the mismatch ratio of 
Fermi velocities.\cite{btk} 
For HFS this formula predicts that N/HFS junctions are in the tunneling limit,
i.e., low transmission,
and AR cannot occur, contrary to experimental observations. Attempts to correct this have been ad-hoc by 
postulating boundary conditions at the interface that are unaffected by the 
mass enhancement of the itinerant heavy electrons.\cite{Deutscher1994}
Alternatively, a single heavy-band tunneling model with an energy-dependent quasiparticle lifetime was proposed
to explain the strongly reduced AR signal, but it lacks to account for the large voltage asymmetry of the conductance
in the normal state.\cite{Gloos1996,Anders1997}

Over time various models for Kondo scattering without magnetic impurities have been proposed to explain point-contact studies
ranging from tunneling into two-level tunneling systems coupled strongly to the conduction electrons \cite{Ralph1992}
to localized electron spins at the point contact.\cite{Gloos2009} 
Very recently, Malteseva et al.\cite{Maltseva2009} presented a theory for electron cotunneling into a dense Kondo lattice that can account for a Fano line shape in the conductance.
Also Yang \cite{Yang2009} argued that the PCS spectra for \cecoin\ are consistent with a two-fluid picture based on the Kondo lattice scenario. 
Common shortcomings of all these approaches have been the ad-hoc nature of additional parameters to explain the PCS spectra,
the neglect of localized states at the interface and the effects of 
pair-breaking surfaces in unconventional superconductors.

In this paper, we present a multichannel tunneling model for the solution of PCS and STS
in heavy-fermion materials that is derived from an analysis of the PCS measurements of the heavy-fermion 
superconductor \cecoin.
While we focus here on \cecoin\, the proposed tunneling model has applications to heavy-fermion systems in general.
For the first time, our multichannel tunneling model quantifies the reduced AR signal and conductance asymmetry observed in 
normal-metal/heavy-fermion superconductor (N/HFS) junctions without special constraints on tunneling barriers, Fermi velocity
mismatch, or itinerant band mass renormalization at the interface.

The article is organized as follows: In Sec.~\ref{framework} we introduce the multichannel tunneling model for the 
HFS and discuss specific limits. In Sec.~\ref{results_and_discussion} we present our theoretical results for a PCS 
junction and compare with several experimental PCS conductance curves varying over a wide range of voltage biases, 
temperature and orientations. 
Finally, we summarize our results in Sec.~\ref{conclusions}.

\section{Tunneling model}
\label{framework}

Our theoretical understanding of the normal-state properties of heavy-fermion materials is based on either the Kondo lattice 
or periodic Anderson model. The Kondo lattice model
describes localized electronic magnetic moments at each lattice site coupled weakly to an itinerant electron band, while the 
periodic Anderson model describes localized $f$-electrons hybridizing with itinerant electrons.\cite{Hewson1993, lohneysen2007}

\subsection{A microscopic model for heavy fermions}

Solving the Kondo lattice or periodic Anderson model is a formidable task
for multiorbital materials.
Instead we model 
the heavy-fermion materials by two bands of itinerant electrons with localized surface states,
which may be caused
by broken $f$-electron bonds at the surface due to the broken translation symmetry,
\bea
\Ham_{HF}&=&\sum_{\alpha;k,\sigma} {\cal E}_\alpha(k) c^\dagger_{\alpha;k\sigma} c_{\alpha;k\sigma}
+ E_0 \sum_{i\sigma} f^\dagger_{i\sigma} f_{i\sigma} .
\eea
The heavy-fermion Hamiltonian $\Ham_{HF}$ represents two bands of itinerant conduction electrons with band index
$\alpha\in\{{\rm light, heavy}\}$ and localized electrons near the surface with site index $i$.
The operators $c^\dagger_{\alpha;k\sigma}$ ($c_{\alpha;k\sigma}$) create (destroy) an itinerant
electron with momentum $k$ and spin $\sigma$ in band $\alpha$, while operators
$f^\dagger_{i\sigma}$ ($f_{i\sigma}$) create (destroy) an $f$ electron at site $i$ with spin $\sigma$.
${\cal E}_\alpha(k)$ are the respective electronic dispersions and $E_0$ is the energy level
of the localized $f$ electrons.

\begin{figure}[b]
\includegraphics[width=0.95\columnwidth,angle=0]{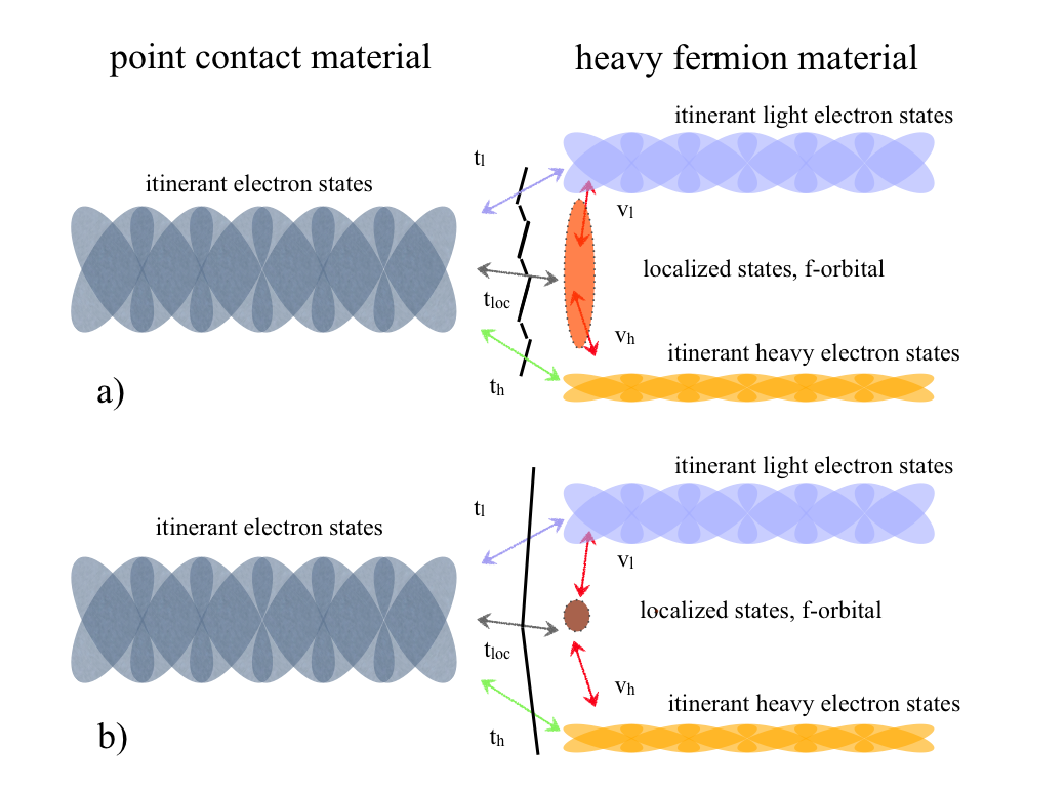}
\caption{(color online) A cartoon of the different tunneling processes from the tip of the point contact to the heavy-fermion
material (localized and itinerant electrons), which are necessary to account for the 
measured asymmetry in point-contact junction conductances and reduced Andreev reflection signals. 
In (a) the localized
state appears as a broad resonance at the interface while in (b) the localized state forms a sharp surface state, 
which acts as a resonant tunneling center. 
}
\label{fig:Setup}
\end{figure}

A simple description of a tunneling experiment is comprised of Hamiltonians for the heavy-fermion material, 
the counter electrode, and the transfer or tunneling processes between them:
$\Ham = \Ham_{\rm HF} + \Ham_{\rm electrode} + \Ham_{\rm T}$.
The counter electrode is given by normal conduction electrons
\be
\Ham_{\rm electrode}=\sum_{k,\sigma} {\cal E}_e(k) e^{\dagger}_{k\sigma}e_{k\sigma}  ,
\ee
and the tunneling Hamiltonian describes all possible transfers
\bea
\Ham_{\rm T} &=& \sum_{\alpha: k,\sigma;k^\prime\sigma^\prime} 
\bigg\lbrack
t^{\alpha}_{k,\sigma;k^\prime\sigma^\prime} e^{\dagger}_{k\sigma} c_{\alpha;k^\prime\sigma^\prime}
+ t^{\alpha}_{k,\sigma;k^\prime\sigma^\prime} c^{\dagger}_{\alpha;k\sigma} e_{k^\prime\sigma^\prime}
\bigg\rbrack
\nonumber\\ &&
+
\sum_{k,\sigma;\sigma^\prime} 
\bigg\lbrack
t^{loc}_{k,\sigma;\sigma^\prime} e^{\dagger}_{k\sigma} f_{i\sigma^\prime}
+
t^{loc}_{k,\sigma;\sigma^\prime} f^{\dagger}_{i\sigma^\prime} e_{k\sigma}
\bigg\rbrack.
\eea
In addition to the standard overlap integrals $t_\alpha$ between the conduction band in 
the point contact and itinerant heavy-fermion bands there is a finite overlap, 
$t_{loc}$, from the point contact to the localized states in the heavy fermion. 
We also account for weak interaction between the localized surface electrons and
itinerant electrons through scattering terms
$v_\alpha$ (see Eq.~7 below).
In general, to get a Fano resonance in the conductance one needs interference between different tunneling 
paths.\cite{fano1961} 
The resulting differential conductance calculated from this model $\Ham_{\rm T}$ will have an asymmetric Fano line shape. 
Our setup is similar to the one for scanning tunneling microscope (STM) tunneling through a magnetic atom on a metallic surface.
\cite{madhavan1998,madhavan2001,Zhao2005,ternes2009} 
The difference between an STM and a point contact is that while the STM is defined by conduction through one or very few quantum channels
a point contact consists of many conducting quantum channels in parallel.
Figure \ref{fig:Setup} shows a cartoon of the processes that are active in tunneling  
between a metallic point contact and heavy-fermion material. 
Here we will extend the picture of conduction through individual quantum channels to a
tunneling model to account for point contacts on a heavy-fermion material. 
We consider strong overlap between electron states in the 
contact and the heavy-fermion compound and thus go
beyond the strict tunneling limit. We then extract microscopic model parameters in form of the overlap integrals
and the energy of the localized state 
from fits made to the asymmetric conductances in the normal state of \cecoin\
reported in the experiments by Goll {\it et al.}\cite{goll2005,goll2006}  and Park {\it et al.}\cite{park2008a,park2008b}.

\subsection{Tunneling current}
To calculate the tunneling current through a quantum channel we employ the standard non-equilibrium Green's function technique.
\cite{Schrieffer1963, caroli1971}
In Keldysh notation the tunneling current {\em per conducting channel} is given by\cite{cuevas2001}
\bea
j(V) &=& \frac{e}{h}{\rm Tr}\, \hat\tau_3 \bigg\lbrack \check t_{loc}\circ \check G_{loc,c}-\check t_{loc}^*\circ \check G_{c,loc}
\nonumber\\ &&
+\sum_{\alpha} \bigg(\check t_\alpha \circ\check G_{\alpha c}-\check t_{\alpha}^*\circ \check G_{c\alpha}\bigg)
\bigg\rbrack^K.
\label{eq:tunnelingcurrent}
\eea
The trace (Tr) is a short-hand notation for summation over momentum and spin $k,\sigma$, and the $\circ$-product indicates a folding over common arguments, e.g., 
$\check t_{loc}\circ \check G_{loc,c}=\sum_{k^{\prime}} t_{loc,\kkp} \check G_{loc,c}(k^\prime,k^{\prime\prime})$,
and $\lbrack\,\,\rbrack^K$ denotes the Keldysh component of the matrix Green's function.
Notation for matrices is: a "hat" ($\hat x$) denotes a Nambu matrix, while a "check" ($\check x$) represents a Keldysh matrix 
(see below).
For ease of readability, we suppress the explicit dependence of the tunneling elements and the Green's function components on momentum
and spin. In equation (\ref{eq:tunnelingcurrent}), $\check G_{i j}$ are Green's function components of the
full matrix in reservoir space ($c$=point contact, $\alpha$= light ($l$) and heavy ($h$) conduction bands of the HF, and
$loc$ is the localized state in the HF). 
It is the components of $\check G$  that straddle the interface  
(e.g. $ \check G_{ch}, \check G_{cl}, \check G_{c,loc}$ etc.) which enter equation (\ref{eq:tunnelingcurrent}).
$\check G$ is to be determined in the presence of a voltage bias across the interface. 
The tunneling Green's function is calculated by summing up single tunneling events in the self-consistent 
non-crossing approximation resulting in the Dyson equation
\bea
\check G &=& \check G^0\!+\!\check G^0\!\circ\! \check V\!\circ\! \check G^0\!+\!\check G^0\!\circ\! \check V\!\circ\! \check G^0\!\circ\! \check V\!\circ\! \check G^0
+\ldots 
\label{eq:tunnelingG}
\eea
or equivalently
\bea
\check G &=& (1 - \check G^0\circ \check V)^{-1} \circ \check G^0 .
\label{eq:tunnelingGfull}
\eea

The transfer matrix $\check V$, which is derived from the tunneling Hamiltonian $\Ham_{\rm T}$, and sketched in
Fig.~\ref{fig:Setup}, details all possible tunneling processes of electrons between the metallic tip, localized and itinerant electrons in the HFS,
\be
\check V_{\kkp}=\left(\begin{array}{cccc} 
0                 & \tilde t_{loc,\kkp}       &  \tilde t_{h,\kkp} & \tilde t_{l,\kkp}\\  
\tilde t^*_{loc,\kkp}  & 0                 &\tilde v_{h,\kkp} & \tilde v_{l,\kkp}\\
\tilde t^*_{h,\kkp} &\tilde v^*_{h,\kkp} & 0              & 0           \\
\tilde t^*_{l,\kkp} &\tilde v^*_{l,\kkp} & 0              & 0           
\end{array}\right) 
.
\ee
Here $\tilde t_{loc,h,l;kk'}$ are the wavefunction overlap integrals between contact and heavy-fermion material making up the
tunneling elements between states $k$ and $k'$, while $\tilde v_{h,l;\kkp}$ are surface-induced scattering elements
between the localized state and the conduction bands $h,l$ in the heavy fermion material.
\cecoin\ is a very pure material with a mean-free path of order of 100 nm.
Therefore, we assume that the junctions are in the ballistic limit where
all tunneling events conserve momentum, i.e.,
$\tilde t_{\alpha,\kkp}=\tilde t_{\alpha}\delta(k-k^\prime)$. 

To further simplify our calculations, we shall assume that the itinerant microscopic Green's functions are described
by quasiclassical Green's functions near the Fermi energy, while it is essential to keep the full energy dependence of
the localized Green's function.
In the case of non-superconducting electrodes the different unperturbed Green's functions are
\bea 
\check G^0_c &=& -i {\cal N}_c  \hat \tau_3\bigg\lbrack \check e^R-\check e^A +2\phi_c(\varepsilon) \check e^K\bigg\rbrack , \\
\check G^0_{h/l} &=& -i {\cal N}_{h/l} \hat \tau_3\bigg\lbrack \check e^R-\check e^A +2\phi_{HF}(\varepsilon) \check e^K \bigg\rbrack  ,
\\
\check G^0_{loc} &=&\eta_{loc}\bigg\lbrack \frac{\check e^R}{\varepsilon^R\hat \tau_3-{E}_0}+ \frac{\check e^A}{\varepsilon^A\hat \tau_3-{E}_0} 
\nonumber\\ 
&+&
\bigg( \frac{1}{\varepsilon^R\hat \tau_3-{E}_0}-\frac{1}{\varepsilon^A\hat \tau_3-{E}_0}\bigg)\phi_{loc}(\varepsilon)\check e^K \bigg\rbrack ,
\label{eq:normalstate}
\eea
with $\varepsilon^{R,A}=\varepsilon\pm i \eta$ 
and $\hat \tau_3$ is the third Pauli matrix in Nambu space.
The distribution functions for the electrons are $\phi_c(T)=\tanh(\varepsilon-eV_{c})/2T,\,
\phi_{HF}(T)=\tanh(\varepsilon-eV_{HF})/2T ,\,\mbox{and}\, \phi_{loc}(T)=\tanh \varepsilon/2T$. 
The parameter $\eta_{loc}$
gives the fraction of localized states and is a dimensionless quantity. 
The 2$\times$2 Keldysh matrix structure of the Green's functions and hopping element above is given as
\bea
&\check e^R=\left(\begin{array}{cc}1&0\\0&0\end{array}\right),\,
\check e^A=\left(\begin{array}{cc}0&0\\0&1\end{array}\right),\,&
\\
&\check e^K=\left(\begin{array}{cc}0&1\\0&0\end{array}\right),\,
\check 1=\left(\begin{array}{cc}1&0\\0&1\end{array}\right). &
\eea

For the Green's function of the localized state, $\check G^0_{loc}$, we assume 
a single level at $\varepsilon={E}_0$. 
To further simplify our model, we will assume that only the heavy electrons undergo a superconducting transition at $T=T_c$, while the light
electrons remain uncondensed. 
Similar arguments were entertained for the unconventional multiband superconductor Sr$_2$RuO$_4$,
\cite{Agterberg1997,Graf2000}
although one might expect, based on general grounds, that both bands go superconducting because
of interband interactions leading to a proximity effect.\cite{Barzykin2007}

The model of a single superconducting band may be justifiable for \cecoin, 
where no multiband gaps have been seen in the PCS data by Park et al.\cite{park2005}
and Goll et al.\cite{goll2006}  However, thermal conductivity measurements have been interpreted 
in terms of uncondensed electrons \cite{Tanatar2005} or fully paired electrons.\cite{Movshovich2001, Seyfarth2008}
In this work, we shall take the view point that the light electrons remain uncondensed and explore the 
theoretical consequences for the PCS spectra.
Thus for temperatures $T<T_c$ only the heavy electron Green's function $\check G^0_h$ becomes
\bea 
\check G^0_h &=&{\cal N}_h
\bigg\lbrack \hat g^R_{surf}(\varepsilon) \check e^R+\hat g^A_{surf}(\varepsilon) \check e^A 
\nonumber\\&&
+ (\hat g^R_{surf}(\varepsilon)-\hat g^A_{surf}(\varepsilon) )\phi_{HF}(\varepsilon) \check e^K \bigg\rbrack\ ,
\label{eq:scstate}
\eea
with $\hat g^{R,A}_{surf}(\varepsilon)$ the self-consistently determined surface Green's functions.
It is worth noting that the surface Green's functions are calculated in the quasiclassical approximation. They can
fully account for surface pair-breaking due to the crystallographic orientation of the surface,
disorder pair-breaking, and realistic band structure.\cite{bruder1990,buchholtz95,lofwander03,fogelstrom04}
Surface pair-breaking can happen for anisotropic order parameters depending on the crystal orientation, 
but not for isotropic s-wave order parameters for which the standard BTK expressions were derived.

The
$\hat g^{R,A}(\varepsilon)$ are the usual (quasiclassical) retarded and advanced superconducting Green's functions.
Here we made the usual quasiclassical approximation of a constant
density of states per unit energy and per spin for conduction electrons at the Fermi surface, 
both in the contact of the metal tip ($\check G^0_c$)
and the heavy fermion ($\check G^0_\alpha$): 
$ {\cal N}_{c,{h,l}}(\varepsilon)=\sum_{k\in BZ} \delta(\varepsilon-{\cal E}_{c,{h,l}}(k))\approx {\cal N}_{c,{h,l}}$. 
However, we kept the energy dependence of the localized density of states.
To compactify our notation, we move the density of states factors ${\cal N}_{c,{h,l}}$ and
$\eta_{loc}$ into the tunneling elements by re-writing 
$\sqrt{\eta_{loc}{\cal N}_c}\,\, \tilde t_{loc} \rightarrow t_{loc}$,
$\sqrt{{\cal N}_c {\cal N}_{h,l}}\,\, \tilde t_{h,l} \rightarrow t_{h,l}$,
$\sqrt{\eta_{loc} {\cal N}_{h,l}}\,\,  \tilde v_{h,l} \rightarrow v_{h,l}$.
The new tunneling elements $t_{h,l}$ are dimensionless,
while $t_{loc}$ and $v_{h,l}$ have dimension $\sqrt{\mbox{Energy}}$.
 
Now we can solve through matrix inversion for the full Green's function $\check{G}$ in 
Eq.~(\ref{eq:tunnelingGfull}). 
With the previous definitions we calculate the Green's function components
needed in the expression for the tunneling current in Eq.~(\ref{eq:tunnelingcurrent}) 
through a point contact. 
First, we focus on the
case of a heavy-fermion material in the normal state. 
For this case, we derive an analytic expression for the 
differential conductance for a single quantum channel, which leads to a modified Fano expression
for the multiband tunneling model
\bea
\frac{dI}{dV}(V) &=& {\cal D} \frac{e^2}{\hbar} \frac{1}{T} 
\, \int^\infty_{-\infty} \frac {d \varepsilon}{2 \pi}
\frac{(q_F \Gamma +\varepsilon-{\tilde E}_0)^2+{\cal B}^2}{\Gamma^2+(\varepsilon-{\tilde E}_0)^2 }\, 
\nonumber\\&& \times
\cosh^{-2}\bigg\lbrack\frac{\varepsilon-eV}{2T}\bigg\rbrack .
\label{eq:normalstatecurrent}
\eea
In Eq.~(\ref{eq:normalstatecurrent}) 
${\cal D}$ is the transparency, ${\tilde E}_0$ is the tunneling-renormalized 
position of the localized energy relative to the Fermi level, $\Gamma$ is the half-width of the resonance,
and $q_F$ is the quantum interference parameter that controls the resonance shape. 
The additional parameter ${\cal B}$
is present for multiband models only, when tunneling through a resonant localized state couples differently
to the HF conduction bands (see below).  ${\cal B}$ adds a Lorentzian to the conventional Fano resonance
which may be absorbed into a complex Fano parameter
$q_F \to q = q_F + i\, q_{\cal B}$ with $q_{{\cal B}}={\cal B}/\Gamma$. 
If one introduces the following two angles $\theta_t$ and $\theta_v$ and writes
\begin{eqnarray}
t_h=t \cos\theta_t ,&& v_h =v \cos\theta_v ,\\ 
t_l=t \sin\theta_t ,&& v_l =v \sin\theta_v ,
\end{eqnarray}
then the five phenomenological model parameters introduced in Eq.~(\ref{eq:normalstatecurrent}) depend on five microscopic 
parameters $(t, v, t_{loc}, E_0,\delta=\theta_v-\theta_t)$ given by the following relations
\begin{eqnarray}
{\cal D}&=&\frac{4 t^2}{(1+ t^2)^2}\label{eq:parameters1}\\
{\tilde E}_0&=&{E}_0-\frac{2 t_{loc} v t \cos\delta}{1+t^2}\label{eq:parameters2}\\
\Gamma&=&\frac{t_{loc}^2+v^2(1 + t^2 \sin^2\delta)}{1+t^2}\label{eq:parameters3}\\
q_F&=&\frac{1}{2}\frac{({\tilde E}_0-{E}_0)}{\Gamma}
\frac{1-t^2}{t^2}\label{eq:parameters4}\\
{\cal B}&=&\frac{v}{t}\sqrt{(t_{loc}^2+v^2 t^2)}\, \sin\delta .
\label{eq:parameters5}
\end{eqnarray}
The angle $\theta_t$ is a free parameter that quantifies the relative weight of the overlap integrals $t_{h,l}$.
$\theta_t$ cannot be determined from normal-state PCS measurements. 
The transparency of a single channel, ${\cal {D}}$, depends only on $t=\sqrt{t_l^2+t_h^2}$. The parameter ${\cal {B}}$
is non-zero only if $\delta\neq 0$, i.e., $\theta_t\neq \theta_v$. 
Finally, we see that in order to observe a Fano-shaped
normal-state conductance, i.e., a finite $q_F$, one needs to be in the limit of 
small to intermediate tunneling coupling,
$t^2 < 1$ or ${\cal {D}}<1$, and have a sizable renormalization due to tunneling into the localized state, 
${\tilde E}_0\neq E_0$.

At this point we like to comment that the differential conductance derived in our multichannel tunneling model with localized
nonmagnetic surface states, shown in Eq.~(\ref{eq:normalstatecurrent}), 
reduces to the Fano expression by Yang\cite{Yang2009} obtained for a Kondo lattice model with hybridized $c$-$f$ electrons. 
Hence, we conclude that the observation of a Fano line shape in the conductance is not a sufficient condition for probing
bulk $f$ electrons in the normal state. In order to associate the Fano peak with bulk $f$ electrons
additional tests are necessary, e.g., observing the bare localized state $E_0$ crossing the Fermi level with temperature or
a magnetic field dependence of the line width.

\begin{figure*}[t!h]
\includegraphics[width=0.8\textwidth,angle=0]{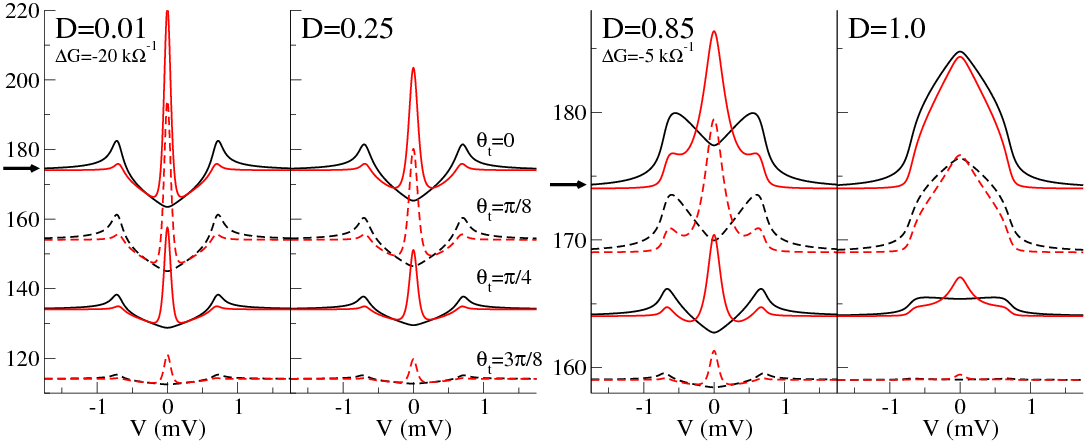}
\caption{(color online) PCS spectra calculated self-consistently for a two-band model in the absence of scattering to localized states and between
bands
(i.e. $t_{loc}=v=0$). We assumed
d-wave superconductivity in one band and uncondensed electrons (normal state) in the other band.
The black and red lines are for a [100] ($0^\circ$) and a [110] ($45^\circ$) interface,
respectively. We vary the transparency ${\cal D}$ from tunneling (left panel) to high transmission (right panel). 
The relative weight of tunneling, $\theta_t$,  into either band is varied from top to bottom for given ${\cal D}$. 
For clarity the conductances for different $\theta_t$ are shifted vertically by $\Delta G$.
The zero-temperature gap is $\Delta_0=$0.6 meV and the background conductance is $G_0=174\, (k\Omega)^{-1}$.
}
\label{fig:dIdV_twoband}
\end{figure*}

The PCS experiments measure a weak Fano-like conductance superimposed on
a large background conductance $G_0(V)$. The asymmetry of the Fano shape in the conductance
with respect to voltage accounts only for 3-5\% of the total conductance. 
To capture both these contributions we need to go beyond the single quantum-channel 
conductance calculated in Eq.~(\ref{eq:normalstatecurrent}). 
Since a point contact forms
over a sizable area, ${\cal {S}}$, compared to atomic scales ($\sim k_F^{-1}$),
one should expect thousands of channels present, each acting as a single quantum channel. 
The conductance should then be written as a sum over the contributions of the individual channels
\begin{equation}
\bigg(\frac{dI}{dV}\bigg)_{PC}=\sum_{i \in channels}  \bigg(\frac{dI}{dV}\bigg)_{i}.
\end{equation}
The interface of a point contact is probably not atomically smooth, which 
means that the majority of localized states near the surface will be broadened to resonances 
due to destructive scattering in the interface (see panel (a) in Fig.~\ref{fig:Setup}). 
In the simplest case, this means that
the localized state is broadened into a resonance as prescribed by
${E}_{0}\rightarrow{E}_{0}+ i \gamma_{broad}$. If the resonance 
is broad enough the conductance kernel in Eq.~(\ref{eq:normalstatecurrent}) 
gives only a broad and featureless contribution resulting in the background conductance. This applies 
to our analysis if $\gamma_{broad}\sim 20-100$ meV is the largest energy scale
of the problem. 
From these types of tunneling channels, i.e., those dominated by $\gamma_{broad}$, 
Eq.~(\ref{eq:normalstatecurrent}) gives a weakly voltage-dependent background conductance 
$G_0(V)\approx G_0+\delta G_0(V)$.   
For the remaining few channels, the localized state forms a sharp surface state with 
${E}_{0}\rightarrow{E}_{0}+ i \gamma_{sharp}$, where
$\gamma_{sharp}$  is much smaller compared to other energies of the problem.     
To describe these channels we extract a phenomenological parameter 
$C_0=n_{sharp} {\cal D} e^2/\hbar$ from the PCS data ($n_{sharp}$ is the fraction of channels having a sharp
localized state at the interface). 
$C_0$ gives the proper magnitude of the Fano resonance relative to the background conductance.
This bimodal distribution of two types of tunneling channels can be resolved in the normal state 
by numerical fits to the various data sets being considered here. 
For the tabulated values of $C_0$ and $G_0$, see Table~I, we find that their ratio ($C_0/G_0$) is typically 
between 3-5\% for the point contacts we analyze. 
Another important point is that both $C_0$ and $G_0$ are only weakly
temperature dependent, which indicates that the bimodal distribution of channels is a stable 
feature of each contact realization in a PCS experiment.
Therefore the phenomenological differential tunneling conductance expression for a point contact accounting for this
type of distribution of single quantum channels becomes
\bea
\frac{dI}{dV}(V) &=& C_0 
\, \int^\infty_{-\infty} \frac {d \varepsilon}{4 T}
\frac{(q_F \Gamma +\varepsilon-{\tilde E}_0)^2+{\cal B}^2}{\Gamma^2+(\varepsilon-{\tilde E}_0)^2 }\, 
\nonumber\\&& \times
\cosh^{-2}\bigg\lbrack\frac{\varepsilon-eV}{2T}\bigg\rbrack
+ G_0(V). 
\label{eq:fit}
\eea
The integrand in equation (\ref{eq:fit}) depends on a set of four parameters $({\tilde E}_0, \Gamma, q_F, {\cal B})$ that
can be extracted from PCS experiments. These determine four microscopic parameters 
$(t_{loc}(t), v(t), E_0(t),\delta(t))$ that depend on the transparency ${\cal D}$ through the tunneling parameter $t$. The 
factors $C_0, G_0(V)$ are determined from the large voltage-scale conductance and in principle determine the distribution of channels $(\gamma_{broad}, \gamma_{sharp})$. 
In what follows we will assume the simplest bimodal distribution of only two
possible values of 
$\gamma_{broad}>\Gamma$ and $\gamma_{sharp}\to 0$.
So far $\theta_t$ is the only model parameter undetermined by normal-state PCS data. 

One should think about the sharp channels as a set of single-channel atomic point contacts. 
If we have a lattice mismatch between the tip and the HFS one may expect that a two-band model, 
as the one proposed here, should show a distribution of the angle $\theta_t$. It seems natural to assume
that the wavefunction of the tip may have different overlaps with each of the itinerant bands 
in the individual atomic point contacts,
but still have a uniform transmission over the whole contact area. A consequence of this will be that different
metallic tips, e.g., Au vs.\ Pt, should result in different overlaps with the itinerant bands of the HFS.
We conclude that a more general formulation of the differential point-contact conductance should be written as
\begin{equation}
\bigg(\frac{dI}{dV}\bigg)_{PC}=\int d \theta_t \rho(\theta_t) \bigg(\frac{dI}{dV}(\theta_t)\bigg)_{sharp}+G_0 ,
\end{equation}  
where $\rho(\theta_t)$ is the distribution of $\theta_t$ and the (dI/dV)$_{sharp}$ 
is the numerically obtained conductance.
Instead of modeling the distribution $\rho(\theta_t)$, we will look at PCS spectra for different values of $\theta_t$.

\begin{figure*}[t!h]
\includegraphics[width=0.75\textwidth,angle=0]{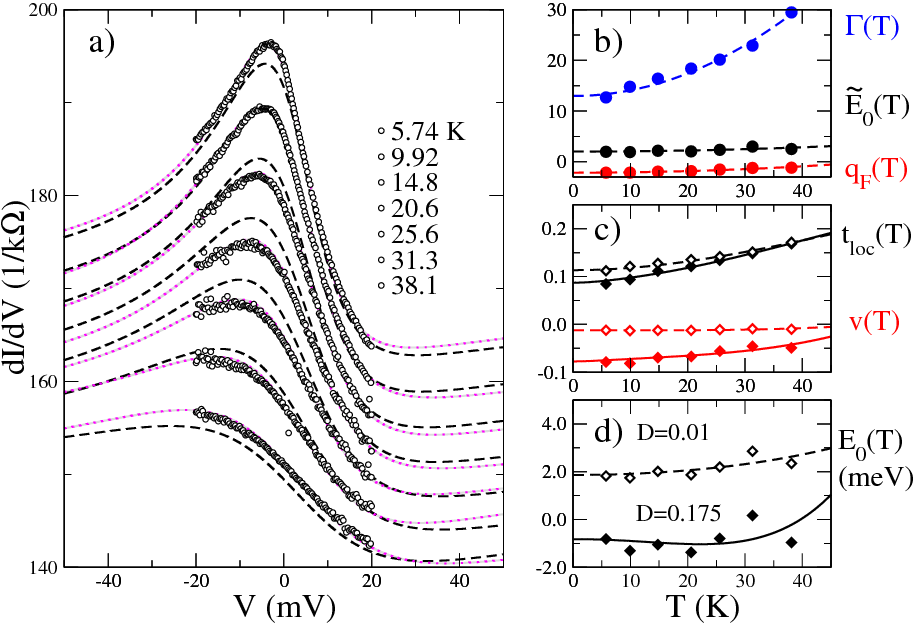}
\caption{(color online) Panel (a): Tunneling fits for several normal-state conductances ($\circ$) taken with a Au-tip on a [001] surface of 
\cecoin\ for a set of temperatures. The curves are shifted down by 4 $({\rm k}\Omega)^{-1}$ for each temperature trace. 
Panels (b)-(d): Fitted Fano parameters and extracted microscopic model parameters based on dI/dV-curves in (a).
The dotted lines in (a) are dI/dV-curves calculated from Eq.~(\ref{eq:tunnelingcurrent}) using 
the microscopic parameters (open diamonds $\Diamond$; closed diamonds are a lower bound for parameters) 
shown in panels (c) and (d), which were extracted from the 
fitted Fano parameters (solid circles $\bullet$) shown in panel (b). 
We fit each model parameter ($\Gamma$ and ${\tilde E}_0$ are in units of meV) in panel (b) to a temperature dependent function
$x(T)=x_0+x_2\cdot(T/45 K)^2$ and re-calculate dI/dV (dashed lines in panel (a)). 
In panels (c) and (d) we show the dependence of the microscopic parameters $(t_{loc},v)$ and  $E_0$ on temperature
for transparencies $0 < {\cal D} \lesssim 0.175$. For transparencies  ${\cal D} > 0.175$ 
the fitting procedure fails and no solutions are found for Eqns.\ (\ref{eq:parameters1})-(\ref{eq:parameters5}).
Here we set ${\cal B}=0$. 
}
\label{fig:dIdV_normalstate}
\end{figure*}
\begin{figure*}[t!h]
\includegraphics[width=0.75\textwidth,angle=0]{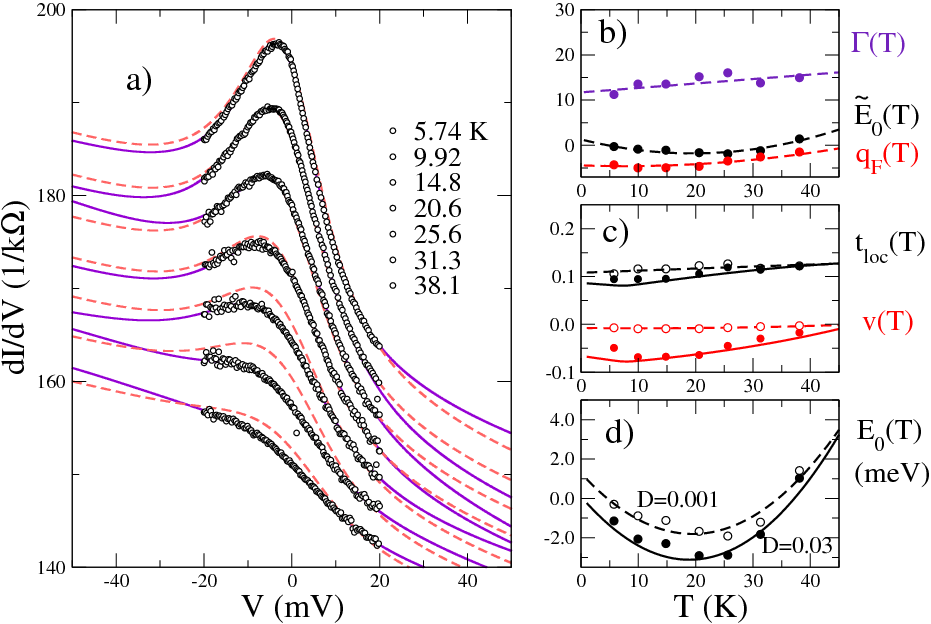}
\caption{(color online) Tunneling fits for the same data as in Fig.~\ref{fig:dIdV_normalstate}, however, assuming a weak
voltage dependent background conductance $\delta G_0(V)$. 
Panel (a): 
The conductance curves are shifted down by 4 $({\rm k}\Omega)^{-1}$ for each temperature trace.
The dotted lines are dI/dV-curves calculated from Eq.~(\ref{eq:tunnelingcurrent}) using 
the microscopic parameters (solid $\bullet$ and open $\circ$ circles) shown in panels (c) and (d), which were extracted from the 
fitted Fano parameters (solid circles $\bullet$) in panel (b). 
Again, we fit each model parameter in panel (b) to a temperature dependent function
$x(T)=x_0+x_1\cdot(T/45 K)+x_2\cdot(T/45 K)^2$
and re-calculate dI/dV (dashed lines in panel (a)). 
In panels (c) and (d) we show the dependence of the microscopic parameters $(t_{loc},v)$ and  $E_0$ on temperature
for transparencies $0 < {\cal D} \lesssim 0.03$. For transparencies  ${\cal D} > 0.03$ 
the fitting procedure fails over the full range of temperatures 
and no solutions for Eqns.\ (\ref{eq:parameters1})-(\ref{eq:parameters5}) are found.
At low temperatures though, one can find solutions for transparencies up to ${\cal D} \lesssim 0.047$.
As in Fig.~\ref{fig:dIdV_normalstate}
the parameter ${\cal B}=0$ and  $\Delta_0=$0.6 meV.
\label{fig:dIdV_normalstate_w_background}
}
\end{figure*}

\subsection{Special limits of the tunneling conductance}

The normal-state differential conductance in Eq.~(\ref{eq:fit}) has several instructive limits 
depending on the particular choice of microscopic model parameters.
We consider the following four cases:
\begin{itemize}
\item[(a)] A one-channel tunneling model: Tunneling into one band only, i.e., 
$t_{loc}=0$ and $\theta_t = 0,\pi/2$, reduces to the standard expression for a single channel contact with a 
transparency ${\cal {D}}=4t^2/(1+t^2)^2$ and a featureless differential conductance.%\cite{Harrison1961} 
The Fano parameters $q_F$, $\Gamma$, and $\tilde{E}_0$ all vanish, thus resulting in no Fano resonance.
When the single band goes superconducting one obtains the standard Andreev conductance for HFS.
This case corresponds to the special limit of $\theta_t=0$ ($t_l=0$) of the two-channel model to be
discussed next.

\item[(b)] A two-channel itinerant tunneling model: Tunneling into both light and heavy bands but keeping $t_{loc}=v=0$ will not generate a Fano resonance in the conductance as seen from Eqns.\ (\ref{eq:parameters1})-(\ref{eq:parameters5}) with
$\Gamma=q_F=\tilde{E}_0=0$. 
In the normal state and for a constant density of states at the Fermi level in light and heavy bands, 
this limit gives a conductance which is constant. When the heavy band goes superconducting, $T < T_c$, 
the two-band model gives an Andreev conductance, which may be reduced in signal 
with respect to the background conductance. 
In Fig.~\ref{fig:dIdV_twoband} we show conductances for fixed values of transparency ${\cal {D}}$,
but with varying relative weight of the tunneling elements via the tunneling angle $\theta_t$. 
The key result of these self-consistent calculations (for details see Sec.~\ref{symmetry}) is that for 
high transparency junctions ($D\to1$) and overwhelmingly tunneling into paired heavy electrons ($\theta_t<\pi/8$) 
it is impossible to differentiate between a $d$-wave superconductor with 
nodal lines along [100] vs.\ [110]. 
Note that for $\theta_t=\pi/4$ the re-normalized tunneling matrix element for paired heavy and normal light 
electrons is equal.
The situation is reversed for $\theta_t=3\pi/8$ when tunneling is predominantly between the normal metallic tip and the
normal light electrons in the HFS, see bottom curves in Fig.~\ref{fig:dIdV_twoband}.
This generic two-band tunneling model demonstrates that in principle
PCS data can differentiate 
between tunneling preferentially into paired heavy electrons versus uncondensed light electrons.

\item[(c)] A two-channel hybridized tunneling model: Simultaneously tunneling into a single itinerant band 
(here we consider a single heavy band, i.e., $t_l=0$ and $t_h\neq 0$) 
and a localized state
$t_{loc}\neq 0$ with nonzero scattering between localized and itinerant heavy electrons, i.e., $v_l=0$ 
and $v_h\neq 0$. 
This point-contact tunneling setup will generate a Fano resonance in the differential conductance 
with $q_{\cal B}={\cal B}=0$, for details see previous general
Eqns.\ (\ref{eq:parameters1})-(\ref{eq:parameters5}).
Very recently, Yang\cite{Yang2009} discussed a Kondo lattice model of hybridized $c$-$f$ electrons.
He derived a normal-state Fano conductance similar to the one in our microscopic model in Eq.~(\ref{eq:normalstatecurrent}).
In order to fit the Fano resonance in \cecoin, 
he introduced  a voltage-dependent Fano parameter $\Gamma(V)$
and a large interband term $q_{\cal B} > q_F$, which he attributed to multiband and correlation
effects beyond $c$-$f$ hybridization of electrons. 
Assuming a constant $q_{\cal B}$ implies a strongly temperature dependent interband scattering
coefficient ${\cal B} \sim \Gamma(T)$, which is difficult to reconcile within our model.
Furthermore, as we have shown above, introducing a nonzero 
$q_{\cal B}$ is equivalent to introducing asymmetric scattering between different itinerant bands and localized states, 
$\delta\neq 0$, invalidating the model assumption of only two species of electrons used in
the standard Kondo lattice model with only $c$-$f$ hybridization.

\item[(d)] A multichannel tunneling model: Simultaneously tunneling into localized and itinerant bands is of topic 
interest and will be discussed in the next section.
\end{itemize}

\section{Results and Discussion}
\label{results_and_discussion}

We take the following approach for extracting the microscopic tunneling elements
$(t,\theta_t,v,t_{loc},\delta,E_0)$ from point-contact conductance data: We fit the differential conductance
in Eq.~(\ref{eq:fit}) to an experimental dI/dV-curve at a given temperature by extracting all
model parameters, i.e., the four Fano parameters
${\tilde E}_0, q_F, \Gamma, {\cal B}$, the relative weight of the Fano-like conductance
$C_0$, and the background conductance $G_0(V)$. 
After extracting the set of phenomenological parameters (${\tilde E}_0, q_F, \Gamma, {\cal B}$)
through numerical fits, we solve for the set of microscopic parameters
($E_0, v, t_{loc}, \delta$) by treating $t$ and $\theta_t$ as free fit parameters.
Consequently, the parameters $t$ or equivalently ${\cal D}=4t^2/(1+t^2)^2$ and $\theta_t$ can only be determined
by studying the conductance in the superconducting state as they effectively drop out from the
normal-state conductance, see Eq.~(\ref{eq:fit}). 

\subsection{Model parameters in the normal state}
In Figs.~\ref{fig:dIdV_normalstate} and \ref{fig:dIdV_normalstate_w_background}, we show results extracted from 
a set of conductances taken at different temperatures using  a Au-tip on $[001]$-oriented \cecoin. For each individual dI/dV-curve we get a good fit to a Fano resonance over the entire measured voltage window of $|V|\leq$ 20 mV. 
We note that while the dI/dV-curves are well fitted within the chosen voltage window in both Figs.~\ref{fig:dIdV_normalstate} and \ref{fig:dIdV_normalstate_w_background}
the extrapolated large voltage-scale conductances are very different as we assumed a constant background $G_0$ in 
Fig.~\ref{fig:dIdV_normalstate}, while in Fig.~\ref{fig:dIdV_normalstate_w_background} we
modeled the background conductance as $G_0(V)=G_0 - G_1 \tanh (V/V^\ast)$, where $V^\ast$ is an additional parameter that is always of the order of the voltage window 
and will not be considered any further.
As seen in these figures, the Fano parameters 
$({\tilde E}_0(T),q_F(T),\Gamma(T),{\cal B})$ depend sensitively on the details of 
how the background conductance is modeled.
Therefore, for deriving meaningful Fano parameters from PCS measurements 
it is very important to measure over voltage biases as large as possible while at the same time avoiding heating.

\begin{figure*}[bt!h]
\includegraphics[width=0.45\textwidth,angle=0]{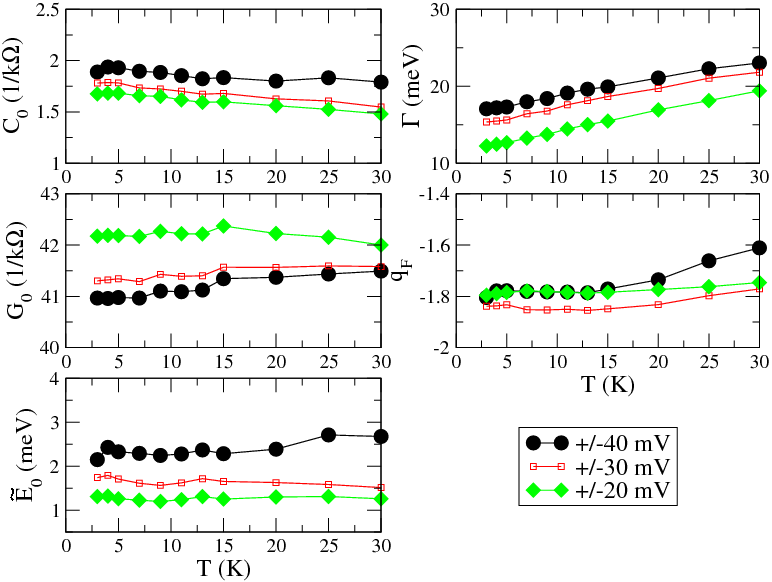}
\hfill
\includegraphics[width=0.45\textwidth,angle=0]{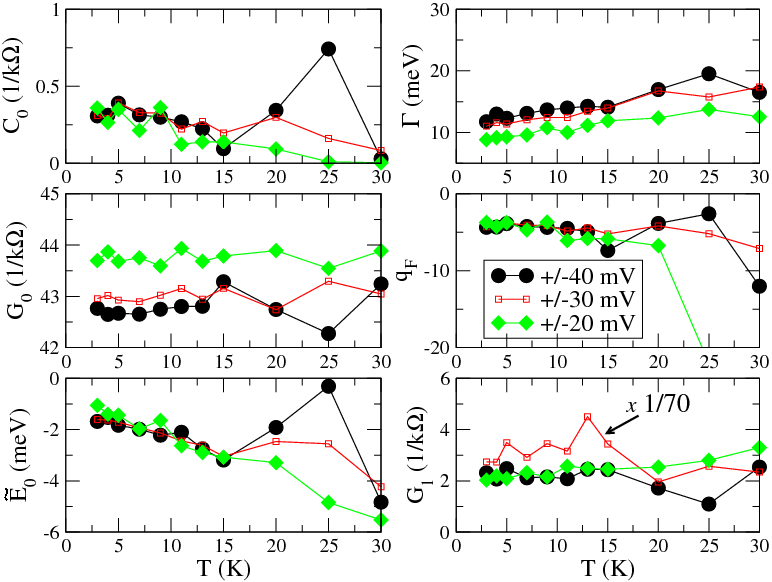}
\caption{(color online) Model parameters vs.\ temperature from fits to the normal-state conductances 
taken with a Pt-tip on a [001] surface of \cecoin\  from 
Ref.~\onlinecite{goll2006}.
The voltage intervals over which fits were performed was varied between 
$|V| < 20$  mV, $|V| < 30$  mV, and $|V| < 40$  mV, 
to test for robustness of fit parameters and fit procedure.
All fits were constrained by setting ${\cal B}=0$.
Left panel: Results for a constant background conductance $G_0(V)=G_0$.
Right panel: Results for a $V$-dependent background conductance $G_0(V)=G_0 - G_1 \tanh (V/V^\ast)$.
}
\label{fig:Fano_parameters_Goll}
\end{figure*}

\begin{figure*}[bt!h]
\vspace*{0.5truecm}
\includegraphics[width=0.35\textwidth,angle=0]{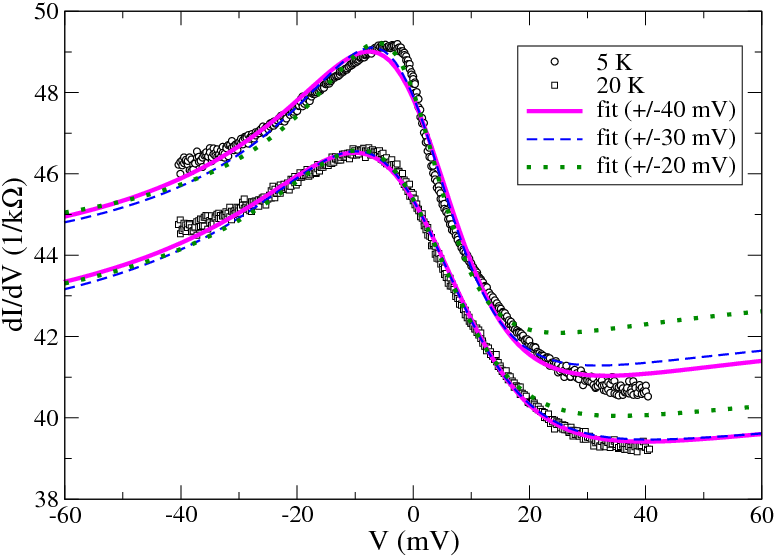}
\hspace*{1cm}
\includegraphics[width=0.35\textwidth,angle=0]{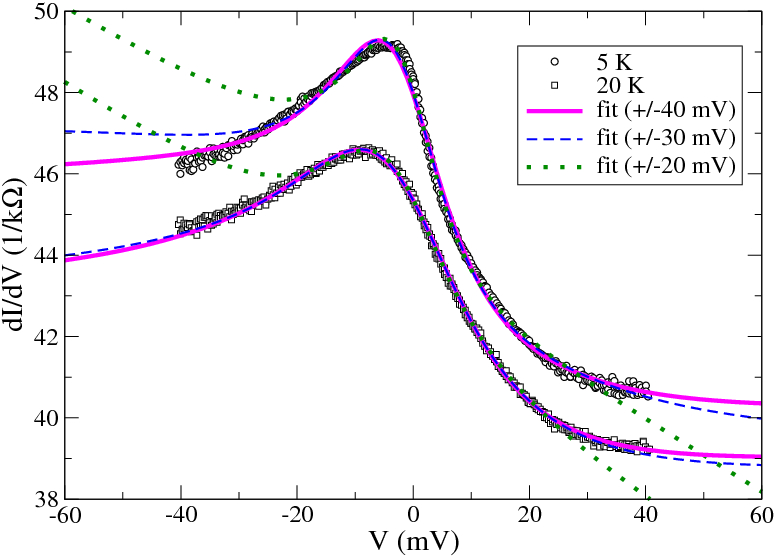}
\caption{(color online) Tunneling fits to the normal-state conductance data (symbols) at 5 K and 20 K with parameters displayed in 
Fig.~\ref{fig:Fano_parameters_Goll}.
For clarity the 20 K data set has been downshifted by 2/k$\Omega$.
Left panel: Assuming $G_0(V)=G_0$, nearly perfect fits are possible in selected voltage windows but not beyond.
Right panel: Assuming $G_0(V)=G_0 - G_1 \tanh (V/V^\ast)$, nearly perfect fits are possible in selected voltage 
windows, as well as outside for large voltage windows in the case of the 20 K data set.
}
\label{fig:Fano_voltage_Goll}
\end{figure*}

\begin{figure*}[ht!b]
\includegraphics[width=0.95\textwidth,angle=0]{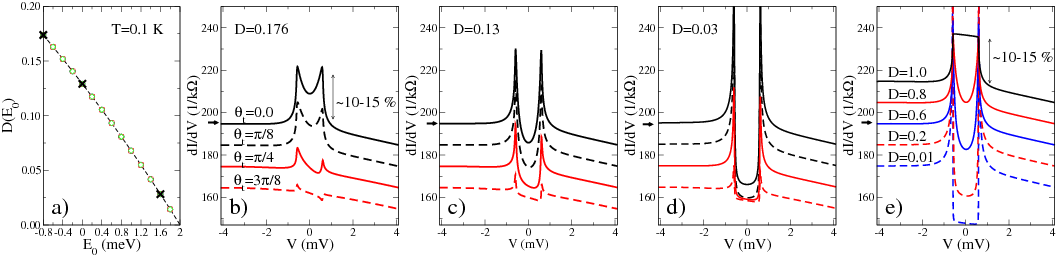}
\caption{(color online) Conductance for an s-wave superconductor with model parameters extrapolated from Fig.~\ref{fig:dIdV_normalstate}.
Panel (a): We plot the transparency ${\cal D}(E_0)$ as a function of the energy relative to the Fermi energy of the localized state at the interface. Panels (b)-(d): The conductance is plotted for a range of tunneling angles $\theta_t$. 
Panel (e): The conductances are compared to the standard BTK conductance vs.\ transparency ${\cal D}$ superimposed on the normal-state conductance with a Fano line shape. We assumed that 15\% of the background conductance is superconducting, i.e, $\eta_{BTK}=0.15$ in Eq.~(\ref{eq:btkdidv}). 
Each conductance curve is shifted up or down in steps
of 10 $({\rm k}\Omega)^{-1}$ relative to the conductance marked by the arrow.  
We used $\Delta_0=$0.6 meV.
}
\label{fig:dIdV_vs_Ef_vs_T}
\end{figure*}
\begin{figure*}[ht!b] 
\includegraphics[width=0.95\textwidth,angle=0]{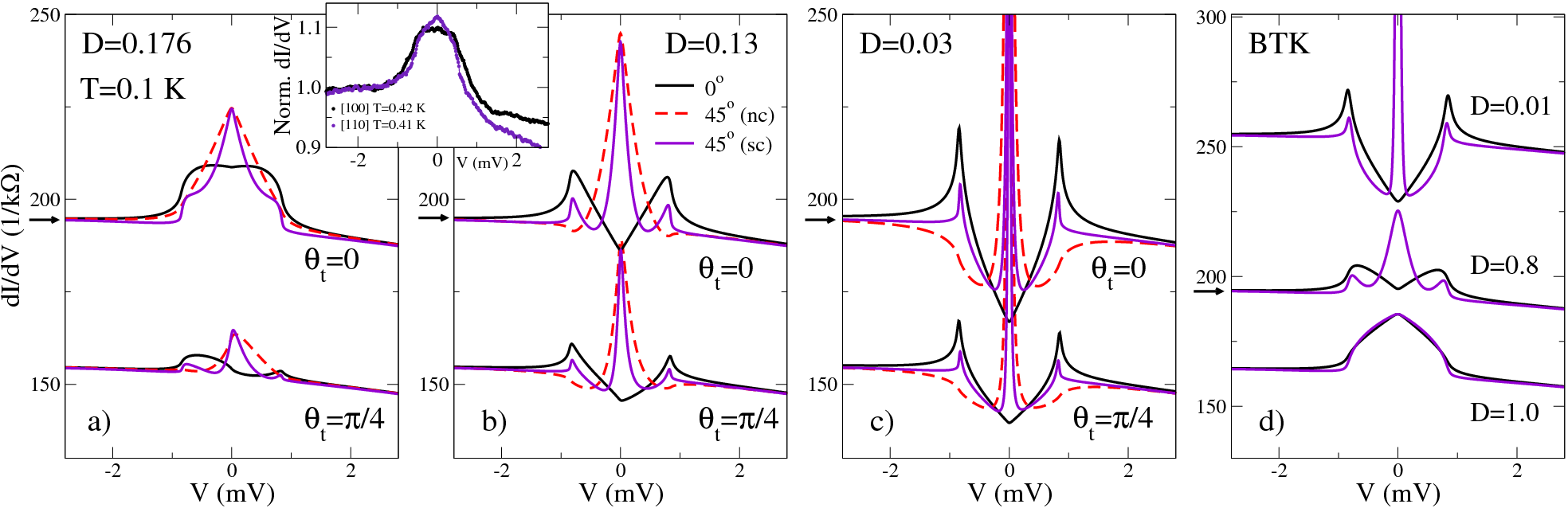}
\caption{(color online) Conductance for a d-wave HFS with the same model parameters as in 
Figs.~\ref{fig:dIdV_normalstate} and \ref{fig:dIdV_vs_Ef_vs_T}. The conductance is plotted for two crystal-to-surface orientations $0^o$ and $45^o$ and for two
values of the tunneling angle $\theta_t$. The labels nc=non-self consistent, sc=self-consistent, refer 
to taking surface pair breaking in to account (sc) or not (nc). 
The conductances are compared with the corresponding self-consistent BTK conductance vs.\ transparency ${\cal D}$ shown 
in panel (d). We assume a superconducting fraction $\eta_{BTK}=0.15$ 
in Eq.~(\ref{eq:btkdidv}).
Each conductance curve is shifted up or down 
relative to the conductance marked by the arrow. 
We used $\Delta_0=$0.6 meV. The inset shows PCS data from 
Ref.~\onlinecite{park2008a}
for comparison. There is a qualitative agreement between the computed PCS in panel (a).
}
\label{fig:dIdV_vs_Ef_dwave}
\end{figure*}
\begin{figure*}[ht!b]
\includegraphics[width=0.95\textwidth,angle=0]{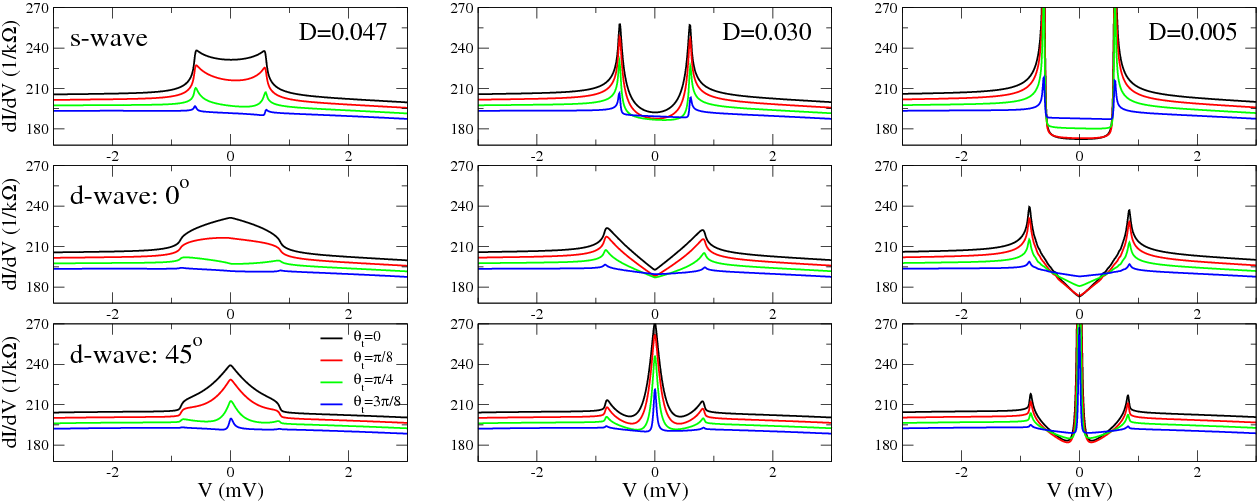}
\caption{(color online) Conductance for an HFS with the same model parameters as in 
Fig.~\ref{fig:dIdV_normalstate_w_background}. The conductance is plotted for both an s-wave and 
a d-wave superconductor with two crystal-to-surface orientations $0^o$ and $45^o$ and for different values of the
tunneling angle $\theta_t$.
We used $\Delta_0=$0.6 meV. The curves have been shifted for clarity.
}
\label{fig:s_n_d_wave_dIdV}
\end{figure*}

The model parameters 
($q_F, {\tilde E}_0, \Gamma, {\cal B},C_0, G_0$) 
all have a temperature dependence that may be fitted with a polynomial $x(T)\approx \sum_{p=0}^{p_{max}} x_p\cdot (T/T^*)^p$, with $p_{max}=1 $ or $2$ and 
$T^*$ is a typical temperature scale for the onset of the strongly correlated heavy-fermion state. 
For \cecoin\ we set $T^*$=45 K. 
When we re-calculate the dI/dV-curves with the fitted temperature dependent 
model parameters, we find that the goodness of a single fit is sensitive to the precise values of 
$q_F, {\tilde E}_0, \Gamma$, and ${\cal B}$. These re-calculated dI/dV-curves are
shown as dashed lines in Figs.~\ref{fig:dIdV_normalstate} and \ref{fig:dIdV_normalstate_w_background}, 
whereas the fits with the original parameter sets are shown as dotted lines.

In Figure \ref{fig:Fano_parameters_Goll} 
we show the temperature dependence of the model parameters, when PCS data are collected over a larger voltage bias window.
We discover that (i) the parameters depend significantly on the size of the voltage bias window over which the fit is performed, 
although they exhibit similar temperature trends;
(ii) the interband scattering parameter ${\cal B}$ is not uniquely determined (not shown).
Only the localized level ${\tilde E}_0$,  Fano line half-width $\Gamma$,  and Fano parameter $q_F$ are insensitive to 
${\cal B}$.
The transparency parameter $C_0$ is weekly dependent on ${\cal B}$. 
Since the parameter ${\cal B}$ has negligible impact on fitting the conductance, we set
${\cal B}=0$ for the remainder of this work.

Quite unexpectedly, we find that the magnitudes and temperature dependences of the  phenomenological model parameters depend strongly on the treatment of the conductance background. The only robust feature that can be extracted is a temperature dependent Fano parameter  $\Gamma$ of order 
16 meV that nearly doubles between 5 K and 30 K. All other parameters depend on the background modeling. 
The extracted temperature dependence of $\Gamma(T)$ suggests the importance of inelastic scattering at elevated temperatures. 
Spin-fluctuations seen in NMR measurements of \cecoin\ and related 
materials are likely sources for this behavior.\cite{Kawasaki2003,Curro2005}
The renormalized
localized level ${\tilde E}_0$ is either positive (above the Fermi level) and nearly independent of temperature (constant background $G_0$) 
or negative (possibly below the Fermi level) and decreases further with increasing temperature ($V$-dependent background). 
We do not observe a crossing of the localized bare
level ${E}_0$ from positive to negative as temperature is increased. This suggests that the localized
states are 
most likely
surface states, since for bulk states localized $f$ electron levels should be pushed below the Fermi
level for temperatures higher than a characteristic heavy-fermion coherence temperature $T^*$. Above $T^*\sim$ 45 K the heavy-fermion system \cecoin\ exhibits well-developed localized 
{$f$} 
moments.

To further characterize the origin of the
resonance, spectroscopic measurements at higher temperatures 
and in magnetic fields may distinguish between localized magnetic Kondo states  in a lattice\cite{Pines2008,Curro2004} 
and nonmagnetic surface states.
Currently the picture is not clear. If the resonance originates from the Kondo lattice effect, one would expect
that the resonance disappears
above the Kondo lattice (coherence) temperature $\sim 45$ K in \cecoin. 
However, this characteristic temperature is only one quarter of what is expected from 
the half-width of the resonance %\cite{madhavan2001,Zhao2005}
$\Gamma\simeq k_B T_K \sim 16$ meV or $T_K \sim 160$ K.
Additionally, it has been suggested that
a magnetic field splits the Kondo resonance due to the Zeeman effect. However, there is no indication that 
fields as high as 9 T affect the conductance.\cite{park09,goll2006b}

In Figure \ref{fig:Fano_voltage_Goll} we show selected normal-state conductance fits
for \cecoin\ at 5 K and 20 K that were used to extract the model parameters shown in Fig.~\ref{fig:Fano_parameters_Goll}. These fits highlight the need for conductance measurements over voltage biases as large as possible, because fitting a Fano resonance
over a small voltage window leads to significant deviations outside that region and hence quite different model parameters.

\begin{table}[b]%[htdp]
\caption{The extracted values of the temperature dependent model parameters from data shown in Figs.~\ref{fig:dIdV_normalstate} and \ref{fig:dIdV_normalstate_w_background}
with constraint ${\cal B}=0$.
}
\begin{center}
\begin{tabular}{|c|c|c|c|c|c|c|}
\hline
expt. & Parameter & ${\tilde E}_0$ & $\Gamma$ & $q_F$ & $C_0$ & $G_0$ \\
\hline
&          & [meV]& [meV] &  & [$({\rm k}\Omega)^{-1}$] &  [$({\rm k}\Omega)^{-1}$] \\
\hline
Fig.~3 & $x_0$                       & 2.01  & 13.0 & -2.16 & 5.6 & 163 \\ 
       & $x_2 \cdot (45\rm{K})^{-2}$ & 1.08  & 22.5 &  1.59 & 2.1 & 2.7 \\
\hline
Fig.~4  & $x_0$                       & 1.30  & 11.7 & -4.44 & 1.66 & 166 \\
        & $x_1 \cdot(45\rm{K})^{-1}$ &-14.2& 4.42 & -1.92 &-3.29 & 3.81 \\
         &$x_2 \cdot (45\rm{K})^{-2}$ & 16.4 & 0 &  5.67 &  4.11 &-0.07\\
\hline
\end{tabular}
\end{center}
\label{default}
\end{table}%

In Table~\ref{default} we report the temperature dependent model parameters for 
data shown in 
Figs.~\ref{fig:dIdV_normalstate} and \ref{fig:dIdV_normalstate_w_background}.
From this analysis we find that the only robust Fano parameter 
is $\Gamma$, which measures the half-width of the Fano resonance,
while all other parameters vary from measurement to measurement and depend on the background fit,
see Figs.~\ref{fig:dIdV_normalstate}, \ref{fig:dIdV_normalstate_w_background} and 
\ref{fig:Fano_parameters_Goll}.
The temperature behavior of the renormalized parameter for the localized level ${\tilde E}_0$ 
can vary from nearly flat to decreasing or increasing with increasing 
temperature depending on the treatment of the background conductance. 
No universal behavior can be identified that might relate to the bulk properties of \cecoin, as 
argued within a two-fluid interpretation of the PCS data. \cite{Pines2008,Yang2009}
How to disentangle surface from bulk effects remains a challenge. It should be possible, 
in principle, to observe the nature of the correlated electronic state 
and the duality\cite{PNAS2008} of $f$ electrons in \cecoin\ with tunneling experiments.

\subsection{Model parameters in the superconducting state}

When a material shows superconductivity one can use the nonlinear voltage dependence of its N/S conductance to extract 
further information about the undetermined parameters ${\cal{D}}$ (or $t$) and $\theta_t$. 
In Fig.~\ref{fig:dIdV_vs_Ef_vs_T}(a) we plot the calculated 
transparency as a function of $E_0$ for the conductances fitted in Fig.~\ref{fig:dIdV_normalstate}. In panels (b)-(d) we show
the corresponding conductances calculated at $T=$0.1 K (\cecoin\ has a $T_c \simeq 2.3$ K) using the extracted
temperature dependent model parameters with an s-wave order parameter. In panel (e) we 
show the self-consistent BTK N/S-conductance \cite{btk} superimposed on the normal-state conductance $(dI/dV)_{Fano}+G_0$ as 
\begin{equation} 
\frac{dI}{dV}=\frac{dI}{dV}_{Fano}+G_0 \bigg\lbrack (1-\eta_{BTK})-
\frac{\eta_{BTK}}{\cal D}\frac{dI}{dV}_{BTK} \bigg\rbrack
\label{eq:btkdidv}
\end{equation}
for the full range of transparencies ${\cal D}$. Here $\eta_{BTK}$ is the fraction of sharp channels
in accordance with the model for PCS described by equation (\ref{eq:fit}).
The self-consistent BTK N/S-conductance, $(dI/dV)_{BTK}$, is computed using quasiclassical Green's functions as done in 
Refs.~\onlinecite{laube00,laube04} 
by accounting explicitly for surface pair-breaking, which goes beyond the standard BTK 
formulation. 
Therefore our self-consistent method
enables the determination of the spectral properties of the superconducting state by
incorporating pair-breaking through
surface scattering for anisotropic order parameters, elastic impurity scattering,
and inelastic scattering off from low-frequency bosonic modes. 

In panels (b)-(d) of Fig.~\ref{fig:dIdV_vs_Ef_vs_T}, we vary the heavy-light electron tunneling 
angle $\theta_t$ allowing for competition
between tunneling into a normal conducting and a superconducting band in the heavy-fermion material. 
As seen, the N/S conductance is sensitive to both the transparency ${\cal{D}}(t)$ and $\theta_t$, i.e.,
the relative ratio of tunneling into the superconducting vs.\ the 
normal-state band. Tunneling through a resonant state enhances the effective transparency of the junction,
so that a ${\cal D}({E_0})\approx 0.17$ has an N/S conductance similar to a BTK conductance of transparency 
${\cal D}\gtrsim 0.8$.
We also see in Figs.~\ref{fig:dIdV_vs_Ef_vs_T} and \ref{fig:dIdV_vs_Ef_dwave}
that for junctions with "high" transparency
${\cal D}({E_0})$, the sub-gap conductance is enhanced.
Another crucial result of this multichannel model is that the conductance
enhancement due to AR is only $\sim 10-15\%$ relative to the normal-state conductance and not the conventional $100\%$. 
Note that the suppression of the AR signal in the HFS comes naturally about by 
tunneling into either multiple bands or through localized states 
into one heavy band. Hence the reduction of the AR signal can be due to ${\tilde E}_0$ not being aligned with the Fermi
level and tunneling is slightly off-resonant. In turn this leads to an incomplete Andreev reflection as
there is a slight particle-hole asymmetry with respect to ${\tilde E}_0$ 
(see $\check G^0_{loc}$ in Eq.~(\ref{eq:normalstate})).    
No ad-hoc postulates are required for the Fermi velocity mismatch between point-contact tip
and HFS or special boundary conditions at the interface.
Instead it is accounted for in the tunneling matrix elements of the wavefunction overlap between the metallic tip
and the HFS.

\subsection{Symmetry of the superconducting order parameter}
\label{symmetry}

There is quite a body of evidence that the heavy-fermion material CeCoIn$_5$ is an unconventional superconductor with
a d-wave symmetry of the order parameter.\cite{Movshovich2001,Izawa2001,Curro2002,Capan2004,Aoki2004,Tanatar2005,Seyfarth2008,park09,An2010}
We follow the strategy outlined in Refs.~\onlinecite{buchholtz95,lofwander03,fogelstrom04,laube00,laube04} and use the quasiclassical theory 
to compute 
self-consistently
the surface states of a d-wave superconductor. The surface Green's function is then used to evaluate the conductance
taking into account surface pair-breaking and hence a reduced order parameter at the surface.
We emphasize that surface pair-breaking of the order parameter is not included in the original BTK formulation and hence will 
lead to differing results.
The results of our self-consistent calculations are shown in Fig.~\ref{fig:dIdV_vs_Ef_dwave} using
the same model parameters as in Fig.~\ref{fig:dIdV_vs_Ef_vs_T} 
As seen in panels (a)-(c) we have a sensitive dependence on the
surface orientation relative to the crystal axis orientation.\cite{tanaka95} 
If the surface normal is aligned with the 
crystal axis
along which the d-wave order parameter $\Delta(T)\, \cos 2(\phi_k-\phi)$ has a lobe
$(\phi=0^o)$, then the conductance shows qualitatively the same shape as an s-wave superconductor, especially if one takes into account 
the trajectory-average over the order parameter $\Delta(T)\, \cos 2(\phi_k-\phi)$. If the surface normal is 
misaligned with the crystal-axis the surface is pair-breaking for a d-wave superconductor as the surface scattering connects 
trajectories with different order parameter values, i.e., 
$\Delta(T)\, \cos 2(\phi_{in}-\phi)\neq\Delta(T)\, \cos 2(\phi_{out}-\phi)$. 
For trajectories where the order parameter changes sign there is a zero-energy Andreev bound state,\cite{Buchholtz1981,Hu94} 
which gives rise to a zero-bias conductance peak.\cite{tanaka95} 
Therefore the NS-conductance in a d-wave superconductor will depend on the principal tunneling direction a point contact or STM tip
has relative to the crystal orientation. In panel (d) in Fig.~\ref{fig:dIdV_vs_Ef_dwave} we show the self-consistent BTK-conductance computed
with Eq.~(\ref{eq:btkdidv}) using the superconducting fraction $\eta_{BTK}=0.15$. 
For ${\cal D}< 1$ there is a clear dependence on the surface-to-crystal orientation.
For fully transparent contacts on the other hand there is very little difference between a 
$\phi=0^o$ contact 
and one with $\phi=45^o$. For the dI/dV-curves calculated using the model parameters 
we never reach the fully transparent limit, as for the
s-wave case in Fig.~\ref{fig:dIdV_vs_Ef_vs_T}, the resonant enhancement of tunneling via a localized state gives for
${\cal D}(E_0)\approx 0.17$ a sub-gap conductance similar to that of the self-consistent BTK scenario with ${\cal D}\gtrsim 0.8$.
For smaller ${\cal D}(E_0)$ the tunneling limit is approached.

The high transmission case, ${\cal D}\to1$, gives good agreement between the self-consistent BTK conductance and 
experimental data\cite{park2008a,park2008b}
at all temperatures below $T_c$ irrespective of surface-to-crystal orientation $\phi$, 
see Fig.~\ref{fig:dIdV_vs_Ef_dwave}.
Similarly, the conductances calculated with the multichannel tunneling model
reproduce to a large extent the experimental PCS data, as can be seen in
Figs.~\ref{fig:dIdV_vs_Ef_dwave} and \ref{fig:s_n_d_wave_dIdV}.
In Fig.~\ref{fig:s_n_d_wave_dIdV} we show the corresponding conductance curves for
an s-wave and d-wave superconductor in the presence of a voltage-dependent
background $G_0(V)$, where the model parameters are extracted form normal-state fits shown in Fig.~\ref{fig:dIdV_normalstate_w_background}.
It is clear from these self-consistent calculations that PCS measurements should be able to differentiate between tunneling into
the nodal vs.\ the antinodal direction of the gap function, irrespective of the transparency of the junction.
Further, in the Fano scenario we find that the amplitude of the conductance in the sub-gap region has the proper suppressed
magnitude compared to the background.
The suppression is a direct consequence of either competing or interfering tunneling channels.    

A key result of these self-consistent calculations is that a modified expression for the BTK conductance 
for an HFS point-contact junction, see Eq.~(\ref{eq:btkdidv}), gives the correct description of experiment.
However, an unphysical parameter ${\cal D}\to 1$, i.e., tunneling in the high transmission limit, is required.
On the other hand, the multichannel tunneling model gives the correct description with 
physically reasonable microscopic parameters, i.e., tunneling in the low transmission limit with ${\cal D}\ll 1$.

\section{Conclusions}
\label{conclusions}

In summary, we developed a 
microscopic 
tunneling model for heavy-fermion materials. 
A narrow spectroscopic feature associated with localized states seen in the heavy-fermion material \cecoin\ is 
modeled through a multichannel tunneling junction. The asymmetric line shape of the differential conductance is understood as a
Fano resonance for localized states in the vicinity of the interface coupled with itinerant heavy electrons in the bulk of \cecoin.

We list the key results of our analysis:
\begin{itemize}
\item The generic two-channel itinerant tunneling model demonstrates that in principle
PCS data can differentiate 
between tunneling preferentially into paired heavy electrons versus uncondensed light electrons.

\item A consequence of the self-consistent calculations for the two-channel itinerant tunneling model
in the absence of localized states
is that for 
high transmission junctions ($D\to1$) and overwhelmingly tunneling into paired heavy electrons ($\theta_t<\pi/8$) 
it is impossible to differentiate between a $d$-wave superconductor with 
nodal lines along [100] vs.\ [110]. 

\item 
The only robust Fano parameter, we succeeded to extract from several PCS measurements on different
samples, is $\Gamma(T)$ with $\Gamma(0)\sim 16$ meV. It measures the half-width of the Fano resonance.
Its temperature behavior suggests the presence of significant inelastic scattering,
which may be due to self-energy effects like scattering off from spin fluctuations or electrons.

\item Finally, for a multichannel tunneling model the zero-bias conductance
enhancement due to Andreev reflection is reduced to only $\sim 10-15\%$ relative to the normal-state conductance 
compared to the conventional 100\% effect. 
The origin of this suppression can be due to either tunneling into multiple competing itinerant bands or through a localized
state into one itinerant band.
\end{itemize}

We conclude that it is desirable to have tunneling measurements 
ranging from the low to high transmission limit. 
Future measurements
at higher magnetic fields, higher temperatures, and over wider voltage bias windows 
will help to probe quantum interference of electrons tunneling
between a metallic tip and itinerant heavy-fermion bands.
In order  to  identify the origin of the localized states
nanosized PCS tips or STM tips, which are in the single-quantum channel limit for tunneling,
will prove to be critical for resolving the mystery of tunneling into
heavy-fermion materials.
Since our microscopic multichannel tunneling model is quite generic, it should also be applicable to other 
heavy-fermion materials.

\section{Acknowledgments}
We are indebted to T. L\"ofwander for insightful discussions early in this work, and especially like to thank 
V. Siderov, J. Thompson, R. Movshovich, and Y. Yang for discussions during later stages of this
work.
M. F.\ was supported by the Swedish Research Council,
W. K. P.\ and L. H. G.\ were supported by the U.S.\ DOE under award No.~DE-FG02-07ER46453 
through FSMRL and CMM at UIUC,
and M. J. G.\ was  supported in part by the U.S.\ DOE at Los Alamos National Laboratory
under contract No.~DE-AC52-06NA25396 and the UC Research Program.

%\input{pcs_hf_prb.bib}
%\input{pcs_hf_prb.bbl}
%\end{document}

\end{document}